\documentclass[showpacs,preprintnumbers,amsmath,amssymb,prd]{revtex4-1}
\usepackage{graphicx}
\usepackage{dcolumn}
\usepackage{bm}
\newcommand{\be}{\begin{equation}}
\newcommand{\ee}{\end{equation}}
\newcommand{\ba}{\begin{eqnarray}}
\newcommand{\ea}{\end{eqnarray}}
\newcommand{\nn}{\nonumber}

\begin{document}

\title{Axial-vector mesons from $\tau\to AP\nu_{\tau}$ decays}

\author{G. Calder\'on}
\email{german.calderon@uadec.edu.mx}
\affiliation{Facultad de Ingenier\'ia Mec\'anica y El\'ectrica,
Universidad Aut\'onoma de Coahuila, C.P. 27276, Torre\'on, Coahuila, M\'exico}
\author{J. H. Mu\~noz}
\email{jhmunoz@ut.edu.co}
\author{C. E. Vera}
\email{cvera@ut.edu.co}
\affiliation{Departamento de F\'isica, Universidad del Tolima A. A. 546, Ibagu\'e, Colombia}

\begin{abstract}
Axial-vector mesons $a_1(1260)$, $f_1(1285)$, $h_1(1170)$, $K_1(1270)$, and $K_1(1400)$ can be produced
in semileptonic $\tau\to A P \nu_{\tau}$ decays, where $P$ stands for the pseudoscalar mesons $\pi$ or $K$.
We calculate the branching ratios based in a meson dominance model. The exclusive channels 
$\tau^-\to a_1(1260)^-\pi^0\nu$, $\tau^-\to a_1(1260)^0\pi^-\nu$, and $\tau^-\to h_1(1170)\pi^-\nu$ turn out to 
be of order ${\cal O}(10^{-3})$, the channel  $\tau^-\to f_1(1285)\pi^-\nu$ of order ${\cal O}(10^{-4})$,
and channels $\tau^-\to K_1(1270)^-\pi^0\nu$, $\tau^-\to K_1(1270)^0\pi^-\nu$, $\tau^-\to K_1(1400)^-\pi^0\nu$, 
and $\tau^-\to K_1(1400)^0\pi^-\nu$ of order ${\cal O}(10^{-6})$. These results indicate that the branching 
ratios could be measured in experiments.
\end{abstract}


\maketitle

\section{Introduction}

Recently the BABAR Collaboration \cite{babar2012} performed measurements of the branching fractions of 
three-prong and five-prong $\tau$ decay modes. Specifically, the branching fraction for the 
$\tau^-\to f_1(1285)\pi^-\nu_{\tau}$ decay is measured,

\be {\cal B}(\tau^-\to f_1(1285)\pi^-\nu_\tau) = (3.86\pm 0.25)\times 10^{-4}, \ee

\noindent where the axial-vector meson $f_1(1285)$ is detected using the $f_1\to 2\pi^+2\pi^-$ and 
$f_1\to \pi^+\pi^-\eta$ modes. We have averaged in quadrature the two branching ratios and added the 
statistical and systematic errors reported by BABAR Collaboration \cite{babar2012}.
This result supersedes a value reported by BABAR Collaboration \cite{babar2005}. Besides, the CLEO 
Collaboration \cite{cleo1997} published a branching ratio of $(5.8^{+1.4}_{-1.3}\pm 1.8)\times 10^{-4}$, 
which is also based on the $f_1\to 2\pi^+2\pi^-$ mode. 

The measurements of $\tau\to AP\nu$ decays can give information about the hadronic matrix elements 
$\langle AP|J_\mu|0\rangle$ at the moderate energy regime of $\tau$ decays. This can be relevant 
in nonleptonic two-body $D$ and $B$ decays, where the annihilation contribution plays a relevant role 
in the branching ratios and CP asymmetries, see Refs. \cite{Dphysics,Bphysics1, Bphysics2, Bphysics3}.

The $\tau\to AP\nu$ decays can contribute to the $\tau$ decay modes with four-pseudoscalar mesons in the final 
state. The exclusive $\tau^-\to f_1(1285)\pi^-\nu$ decay contributes as an intermediate state in $\tau$ 
decay involving five-pseudoscalar mesons. The former decays can be related to $e^+e^-\to (4\pi)^0$. 
The relation between the $\tau^-\to (4\pi)^-\nu$ decays and the electromagnetic annihilation 
$e^+e^-\to (4\pi)^0$ process appears because the components of the same current are present in both
cases, i.e., the current $\bar d\gamma_\mu u$ in $\tau$ decay and 
$1/\sqrt{2}(\bar u\gamma_\mu u-\bar d\gamma_\mu d)$ in the case of $e^+e^-$ annihilation. Thus, these two
processes are related by isospin symmetry.

On the other hand, at the theoretical level, the production of p-wave mesons in semileptonic $\tau$ decays has 
not been widely studied in the literature. In effect, Ref. \cite{herman2011} and this paper are the only 
works that  have considered the production of orbitally excited mesons in $\tau \to M P \nu_\tau$ channels, 
where $M$ can be an axial-vector meson or a tensor meson. The exception has been Ref. \cite{li1996}, where
the branching ratio for the channel $\tau^-\to f_1(1285)\pi^-\nu$ estimated is $2.91\times 10^{-4}$.

In this paper we estimate the branching ratios of the $\tau\to AP\nu$ decays, where $P$ is the pseudoscalar meson 
$\pi$ or $K$ and $A$ is the axial-vector meson $a_1(1260)$, $f_1(1285)$, $h_1(1170)$, $K_1(1270)$, or $K_1(1400)$.
We have considered the kinematically allowed, the $G$-parity conserved, and the first-class currents decays.
We use a meson dominance model, where the resonances that contribute are coming from the vector mesons 
$\rho(770)$, $\rho(1450)$, and $K^*(982)$ and the axial-vector meson $a_1(1260)$. The weak decay and strong 
coupling constants are determined from experimental data. 

In our estimations, we find that the branching ratios for the exclusive channels $\tau^-\to a_1(1260)^-\pi^0\nu$, 
$\tau^-\to a_1(1260)^0\pi^-\nu$, and $\tau^-\to h_1(1170)\pi^-\nu$ are of order 
${\cal O}(10^{-3})$. The channels $\tau\to K_1(1270)\pi\nu$ and $\tau\to K_1(1400)\pi\nu$ decays are of order 
${\cal O}(10^{-6})$. Finally, the modes $\tau\to a_1(1260)K\nu$ and $\tau\to K_1(1270)K\nu$ are kinematically 
suppressed. Up to now, the $\tau^-\to f_1(1285)\pi^-\nu$ decay of order ${\cal O}(10^{-4})$ is the only mode 
that can be compared with experimental measurements.

The paper is organized as follows. In Sec. II, we describe the amplitude in the meson dominance model and give 
the expression to estimate the width decay. The calculation of the weak decay and strong coupling constants 
are given in Sec. III. In Sec. IV, we present the estimation of the branching ratios for the modes considered 
in this paper. And finally, we conclude in Sec. V.

\section{Amplitude and width decay for semileptonic $\tau\to AP\nu$ decays}

The decay amplitude for the $\tau(p_\tau)\to A(p_A)P(p)\nu_\tau(p_\nu)$ process, where $A$ 
denotes the axial-vector meson $a_1(1260)$, $f_1(1285)$, $h_1(1170)$, $K_1(1270)$, or 
$K_1(1400)$, and $P$ is the pseudoscalar meson $\pi$ or $K$, can be written by

\be
{\cal M}(\tau \to A P \nu_{\tau}) =\frac{G_{F}}{\sqrt{2}}V_{ui}
\bar{u}(p_{\tau})\gamma^{\mu }(1-\gamma _{5})u(p_{\nu})
\langle A(p_A)P(p)|J_{\mu }(0)|0\rangle \ ,
\ee

\noindent where $G_F$ is the Fermi constant, $V_{ui}$ (i=d or s) is the corresponding element of 
the Cabibbo-Kobayashi-Maskawa matrix, and $J_\mu(0)$ is the $(V-A)$ weak current. 

The hadronic matrix element $\langle A(p_A,\varepsilon)P(p) |J_{\mu}|0\rangle$ is parametrized by \cite{isgw}

\ba \langle A(p_A,\varepsilon) P(p)|J_{\mu}|0\rangle &=& l(t) \epsilon_\mu 
+(\epsilon.p)[c_+(t)(p-p_A)_\mu + c_-(t)(p+p_A)_\mu]\nn\\
&+&iq(t) \epsilon_{\mu\nu\rho\sigma}\epsilon^\nu(p_A-p)^\rho(p_A+p)^\sigma\ \label{hc},
\ea

\noindent where the hadronic matrix element is expressed in terms of the form factors $l(t)$, $c_+(t)$, 
$c_-(t)$, and $q(t)$, which are Lorentz-invariant functions of squared momentum transfer $t=(p_A+p)^2$;
the polarization vector $\epsilon_\mu$ describes the axial-vector meson $A$.  

To estimate the form factors, we use the meson dominance model, i.e., the decay amplitude is given 
by the sum of intermediate meson resonance contributions,

\be 
{\cal M}(\tau \to A P \nu_{\tau}) = 
\sum_{R^*} {\cal M}(\tau \to R^*\ \nu_{\tau} \to A P \nu_{\tau}) \ ,
\ee

\noindent where $R^*$ is the intermediate resonance, which has the right quantum numbers for the 
specific processes. The sum is extended over all possible resonance contributions; see Fig. 1.

The production of the resonance $R^*$ can be classified as either first- or second-class current
depending on the spin $J$, parity P, and G-parity of the resonance particle. In the Standard Model, the
first-class current is considered to dominate. The second-class currents are associated with a decay 
constant proportional to the mass difference between an up and down quark, and in the exact isospin limit 
symmetry they vanish \cite{weinberg58}. The vector meson resonances $\rho(770)$, $\rho(1450)$, $K^*(982)$
and the axial-vector meson $a_1(1260)$ are produced by first-class currents, which are the contributions 
included in this work. The vertex $R^*\to AP$ decay, which is produced by strong interaction, must 
conserve G-parity.

Comparing the hadronic current, see Eq. (\ref{hc}), with the amplitude build from the meson dominance
model, which is calculated using Feynman rules, we determine expressions for the form factors $q(t)=0$ 
and $c_-(t)=-c_+(t)=c(t)$. Thus, the amplitude is expressed in terms of only two form factors $l(t)$
and $c(t)$. These form factors can be expressed by means of the Breit-Wigner (BW) function,

\ba l(t) &=& -{i f_J g_{AJP}\over M_J}\ (p_A\cdot p_J)\ BW_J(t), \nn\\
c(t) &=& {i f_J g_{AJP} \over 2M_J}\ BW_J(t),
\ea

\noindent where the subindex $J = V, A'$ stands for vector or axial-vector meson contributions, i.e.,
$V=\rho(770)$, $\rho(1450)$ or $K^*(982)$ and $A'=a_1(1260)$. The function $BW_J(t)$ is defined by

\be BW_J(t) = {M_J^2\over M_J^2-t-i\sqrt{t}\, \Gamma_J(t)}\ ,\ee

\noindent the parameter $M_J$ is the mass, and the function $\Gamma_J(t)$ is the off-shell width decay 
of the intermediate meson resonant particle, see Refs. \cite{Kuhn1990, Dumm2009}, where they 
established the relevance of considering the moment transfer dependency on the width decay of the resonance
for semileptonic $\tau$ decays. The function $BW_J(t)$ appears from the Feynman rule propagator for
the unstable charged vector or axial-vector mesons considered as intermediate virtual resonance,
which is derived in general form in Ref. \cite{LopezCastro1991},

\ba D^{\mu\nu}_J(q^2) &=& {1\over M_J^2-q^2-i\sqrt{q^2}\Gamma_J(q^2)}
\left[-g^{\mu\nu}+{q^\mu q^\nu \over M_J^2-i\sqrt{q^2}\Gamma_J(q^2)}\right] ,\ea

\noindent where $q^2=t$.

The unpolarized squared amplitude in terms of the form factors $l(t)$ and $c(t)$ is the following:

\ba \sum_{pol} |{\cal \bar M}|^2 &=& 2 G_F^2 |V_{ui}|^2 \Big[a_1(t,s) |l(t)|^2 + a_2(t,s) |c(t)|^2 
+ a_3(t,s) Re[l^*(t) c(t)]\Big] ,\label{amplitude2}
\ea

\noindent where the kinematic factors are defined by

\ba a_1(t,s) &=& {1\over m^2_A} \Big[(s-m^2_\tau)(s+t-m^2_\tau-m^2) + 
m^2_A(m^2+2m^2_\tau-2t-s)\Big] ,\nn\\
a_2(t,s) &=& {1\over m^2_A} \Big[( (m^2-t)^2-2(m^2+t)m^2_A+m^4_A )
( (m^2-s)m^2_A+(s-m^2_\tau)(s+t-m^2_\tau-m^2) )\Big], \nn\\
a_3(t,s) &=& {2\over m^2_A} \Big[(m^2-s)m^4_A + (m^2-t)(s-m^2_\tau)(m^2+m^2_\tau-s-t)\nn\\
&+& m^2_A(s^2 + m^2t + m^2_\tau t - m^4 - m^2m^2_\tau - m^4_\tau)  \Big].
\ea

The kinematics factors $a_1(t,s)$, $a_2(t,s)$, and $a_3(t,s)$ are functions of squared moments 
$t=(p_A+p)^2$ and $s=(p_\nu+p_A)^2$ and the masses of $\tau$ lepton $m_\tau$, axial-vector meson $m_A$,
and pseudo-scalar meson $m$. 

The width decay for the channel $\tau\to AP\nu_\tau$ is calculated integrating the unpolarized amplitude,
Eq. (\ref{amplitude2}), with respect to the variables $t$ and $s$ and the appropriate kinematic limits.

The vector meson resonant contribution is the only contribution for the channels studied in this paper, with
the exception of the $\tau^- \to f_1(1285)\pi^-\nu$ decay, where it has an axial-vector meson $a_1(1260)$
as the only present contribution. Specifically, for the channels $\tau\to a_1(1260)\pi\nu$, 
$\tau\to h_1(1170)\pi\nu$, $\tau\to K_1(1270)K\nu$, and $\tau\to K_1(1400)K\nu$, we have the contributions
of vector meson $\rho(770)$ and $\rho(1450)$ resonances. The strange vector meson $K^*(982)$ contributes 
to the processes $\tau\to a_1(1260)K \nu$, $\tau\to K_1(1270)\pi\nu$, and $\tau\to K_1(1400)\pi\nu$.

In studies of the $\tau^-\to \pi^-\pi^0\nu$ channel carried out by the Belle \cite{Belle2008} and 
ALEPH \cite{ALEPH2005} Collaborations, it was established that in order to correctly describe the spectral 
function, it is necessary to introduce two resonances -the vector mesons $\rho(770)$ and $\rho(1450)$. 
We model the two contributions with a linear combination normalized by 

\be BW_V(t) = {BW_\rho(t)+\beta BW_{\rho'}(t)\over 1+\beta} ,\ee

\noindent where the parameter $\beta$ is determined by a fit to the hadronic spectral function obtained 
by the experiments. In our work, we model in the same manner the contribution of the two resonances 
$\rho(770)$ and $\rho(1450)$ to the channels considered above. We use a conservative value of 
$\beta=\pm(0.2\pm 0.1)$.

\begin{figure}
\includegraphics[width=7cm]{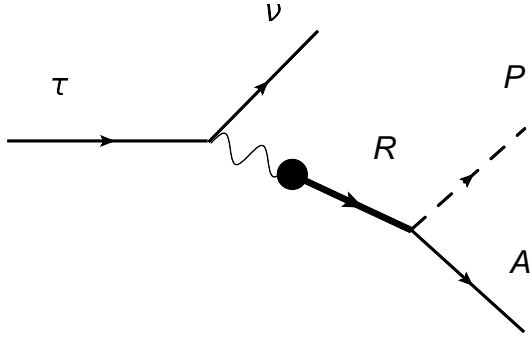}
\caption{Intermediate meson dominance resonance of $\tau\to AP\nu$ decays.}\label{fig}
\end{figure}

\section{Weak decay and strong coupling constants}

In order to obtain the weak decay and strong coupling constants that are required in this work, we make 
use of the available experimental information. However, for some channels, we use isospin $SU(2)$ 
symmetries to relate them to the constants obtained by experimental data.

We use the experimental reported branching ratios in Ref. \cite{pdg2012} to determine the strong 
coupling constants from the decays of strange axial-vector mesons $K_1(1270)\to K\rho$, 
$K_1(1270)\to K^*(982)\pi$, and $K_1(1400)\to K^*(982)\pi$ and from the $h_1(1170)\to \rho\pi$ decay. 

The strong coupling constant $g_{AVP}$ (in $GeV^{-1}$) is determined from the amplitude of probability for the 
$A\to VP$ process, which is defined by

\be {\cal M}(A(p_A,\epsilon_A)\to V(p_V,\epsilon_V) P(p)) = g_{AVP}\epsilon_A^\mu \epsilon_V^\nu
[(p_A.p_V)g_{\mu\nu}-(p_A)_\nu(p_V)_\mu] .
\ee

The decay rate in the rest frame of the decaying axial-vector meson is

\be \Gamma(A\to VP) = {1\over 3}{g^2_{AVP} \over 32\pi}{\lambda^{1/2}(m^2_A,m^2_V,m^2)\over m^3_A} 
\left [2m^2_A m^2_V + (m^2_A+m^2_V-m^2)^2\right] ,
\ee

\noindent where $\lambda(x,y,z)$ is the Kallen function, which is defined by 
$\lambda(x,y,z)=x^2+y^2+z^2-2(xy+xz+yz)$.

Using the measured width decay for the channels $K_1(1270)\to K\rho$, $K_1(1270)\to K^*\pi$, 
$K_1(1400)\to K\rho$, $K_1(1400)\to K^*\pi$, and using the isospin $SU(2)$ symmetry to relate with the charged 
modes, we obtain the following values in ($GeV^{-1}$): $g_{K_1(1270)^- K^0 \rho^-} =5.04 \pm 0.67$, 
$g_{K_1(1270)^0 K^- \rho^-} =5.04 \pm 0.67$,
$g_{K_1(1270)^- K^{*-} \pi^0} =0.69 \pm 0.13$, $g_{K_1(1270)^0 K^{*-} \pi^-} =0.97 \pm 0.19$,
$g_{K_1(1400)^- K^0 \rho^-} =0.68 \pm 0.34$, $g_{K_1(1400)^0 K^- \rho^-} =0.68 \pm 0.34$,
$g_{K_1(1400)^- K^{*-} \pi^0} =1.95 \pm 0.10$, and $g_{K_1(1400)^0 K^{*-} \pi^-} =2.75 \pm 0.14$. 
The error reported is due to the experimental uncertainty in the width decay.

In order to extract the strong coupling constant $g_{h_1\rho\pi}$, we have assumed that only the mode
$h_1(1170)\to\rho\pi$ contributes to the total width decay of the axial-vector meson $h_1(1170)$. We 
obtain $g_{h_1 \rho^- \pi^+} = 3.85\pm 0.21$ GeV$^{-1}$. The uncertainty is determined by the experimental 
error in the width decay. 

To obtain the strong coupling constants $g_{a_1 \rho \pi}$ and $g_{a_1 K^* K}$, we use 
the branching ratios reported in Ref. \cite{pdg2012} for the semileptonic 
$\tau^-\to \pi^-\pi^+\pi^-\nu$ and $\tau^-\to K^-\pi^-K^+\nu$ decays, respectively.
We consider the approximation that the channel $\tau^-\to \pi^-\pi^+\pi^-\nu$ has the 
only contribution from $\tau^-\to a^-_1(1260)\nu\to \rho^0\pi^-\nu$. Using the value
for the weak decay constant $f_{a_1}$ as is discussed below and the meson
dominance model for the process $\tau^-\to \rho^0\pi^-\nu$, we obtain 
$g_{a_1 \rho \pi} = 5.14\pm 2.00$ GeV$^{-1}$, where the principal error is coming 
from the width and weak decay constant of the axial-vector meson $a_1(1260)$. 
The channel $\tau^-\to K^-\pi^-K^+\nu$ has contributions due to vector and axial-vector 
currents. There are different experimental determination of the axial-vector contribution 
which are inconsistent, see Refs. \cite{Davier2005, Davier2008}. These authors consider the
value $(75\pm 25)\%$ to account for the experimental discrepancy. Additionally, in estimates 
such as $\approx 37\%$ \cite{Roig2007} 
obtained by the resonance chiral perturbation theory, the axial-vector contribution is not dominant. 
In view of these considerations we make a conservative assessment for the axial-vector contribution 
of $(65\pm 25)\%$. We approximate the channel $\tau^-\to K^-\pi^-K^+\nu$ to $\tau^-\to K^{*0}K^-\nu$ 
with the only contribution coming in the axial-vector current from the axial-vector $a_1(1260)$. 
We obtain $g_{a_1 K^* K}=15.7\pm 6.7$ GeV$^{-1}$, where the error is principally due to the measured 
axial-vector contribution and parameters associated with the meson $a_1(1260)$, the width, 
and the weak decay constant.

The strong coupling constant $g_{f_1 a_1 \pi}$ can be obtained from the branching ratio of the 
channel $f_1(1285)\to \rho^0\pi^-\pi^+$ reported in Ref. \cite{pdg2012}. This decay can
be modeled by the contribution of the axial-vector meson $a_1(1260)$, given the sequence 
decays $f_1(1285)\to a_1(1260)\pi^+\to \rho^0\pi^-\pi^+$, in the following manner.
Two vertexes are produced by the strong interactions -the first one $f_1(1285)\to a_1(1260)\pi$, and
after that the meson $a_1(1260)$ propagates and decays by the vertex  $a_1(1260)\to \rho\pi$.
Thus, with a value for the strong coupling constant $g_{a_1 \rho \pi}$, we obtain the other
coupling  $g_{f_1 a_1 \pi}=0.73\pm 0.29$ GeV$^{-1}$, where the error is principally coming
from the error in width decay of the intermediate axial-vector meson $a_1(1260)$, the 
strong coupling constant $g_{a_1 \rho \pi}$, and the branching ratio for the channel 
$f_1(1285)\to \rho^0\pi^-\pi^+$.

We have estimated the weak decay constant of hadron mesons $\rho(770)$ and $K^*(982)$ from 
the experimental data on branching ratio from $\tau^-\to h^-\nu_{\tau}$ decays, 
where $h=\rho(770)$ or $K^*(982)$ . We obtain $f_{\rho}=218\pm 2$ and $f_{K^*}=210\pm 5$ MeV. 
In the calculations above, the error in the weak decay constants is coming from the 
experimental error in branching ratios of $\tau^-\to h^-\nu_{\tau}$ decays. 
The weak decay constant for the axial-vector meson $a_1(1260)$ used is $f_{a_1}=165\pm 34$ MeV, 
which is obtained using the value of the peak of the axial-vector meson $a_1$ in the tau
spectral distributions of the decays $\tau^-\to \pi^-\pi^0\pi^0 \nu_\tau$ and 
$\tau^-\to \pi^-\pi^+\pi^- \nu_\tau$ from ALEPH Collaboration \cite{ALEPH2005}. See publicity 
accessible at the web site \cite{ALEPHweb}.

The necessary values for the relevant parameters such as width decays and masses of the 
particles involved in the channels calculated in this work have been taken from 
Ref. \cite{pdg2012}. In particular, for the axial-vector meson $a_1(1260)$, Ref. 
\cite{pdg2012} quotes $250$ to $600$ Mev for its width decay, which in our calculations
we consider $425\pm 175$ MeV.

\section{Branching ratios}

The branching ratios for the exclusive $\tau^-\to AP\nu_\tau$ decays estimated in this work are 
shown in Table I.

The largest branching ratios estimated in this work turn out to be of order ${\cal O}(10^{-3})$, corresponding 
to the exclusive modes $\tau^-\to a_1(1260)^-\pi^0\nu$, $\tau^-\to a_1(1260)^0\pi^-\nu$, and 
$\tau^-\to h_1(1170)\pi^-\nu$. In these modes we have the contribution of two vector meson resonances $\rho(770)$ 
and $\rho(1450)$, which produce broad form factors $l(t)$ and $c(t)$ as a function of the momentum transferred 
$t$, consequently resulting in a large branching ratio estimation.

Up to now, the only measured branching ratio from the calculations in this work is for the channel 
$\tau^-\to f_1(1285)\pi^-\nu$. Our estimate for this channel is $(1.3\pm 1.2)\times 10^{-4}$. We can compare with 
the recent result of BABAR Collaboration $(3.86\pm 0.25)\times 10^{-4}$ \cite{babar2012} and determine that our
estimation differs by $2.1$ sigma error with respect to the experimental result.

The channels with branching ratios of order ${\cal O}(10^{-6})$ are $\tau\to K_1(1270)\pi\nu$ and
$\tau\to K_1(1400)\pi\nu$ decays. They have only the contribution of the vector meson $K^*(982)$ as 
intermediate resonance. 

The exclusive channels $\tau^-\to K_1(1400)^0K^-\nu$ and $\tau^-\to K_1(1400)^-K^0\nu$ decays are not 
kinematically allowed. The modes $\tau\to a_1(1260)K\nu$ and $\tau\to K_1(1270)K\nu$ decays are of the 
order of ${\cal O}(10^{-7})$ and ${\cal O}(10^{-9})$, respectively. This is due to the small phase space 
available for their decay; i.e, they are kinematically suppressed decays.

The channels $\tau\to a_1(1260)\pi\nu$, $\tau^-\to h_1(1170)\pi^-\nu$, $\tau\to K_1(1270)\pi\nu$, and 
$\tau\to K_1(1400)\pi\nu$  decays, with resonance contributions coming from the vector mesons $\rho(770)$ and 
$\rho(1450)$, have the largest error due to the error in the parameter $\beta$. In the other cases, where the 
meson resonance contribution is the strange vector meson $K^*(982)$ or the axial-vector meson $a_1(1260)$, the 
principal error in the branching ratio estimate is due to the uncertainty in the strong coupling constant 
$g_{AJP}$, where $J$ is the resonance contribution $K^*(982)$ or $a_1(1260)$.

In addition, there are two other sources of error in our estimations not considered. We use isospin $SU(2)$ 
symmetry to relate the strong coupling constant to that obtained by experimental information. The error in 
isospin $SU(2)$ symmetry can reach $5\%$ in the estimation. The other source of theoretical uncertainty is 
coming from the contact term in the weak vertex of the hadronic current $\langle AP|J_\mu(0)|0\rangle$. 
However, we do not consider this contribution because it is associated with the continuum, i.e., there is no 
function $BW_V(t)$. The contact term contribution is not possible to estimate in a meson dominance model.

\begin{table}[ht]
{\small TABLE I.~Branching ratios for $\tau\to AP\nu$ decays.}
\begin{center}
\begin{tabular}{l r r}
\hline \hline
AP mode        & Branching ratio  &  Parameter $\beta$ \\
\hline
$a_1(1260)^-\pi^0$ & $(6.9\pm 6.3)\times 10^{-3}$ & $\beta=+0.2\pm 0.1$  \\
$a_1(1260)^-\pi^0$ & $(6.1\pm 5.9)\times 10^{-3}$ & $\beta=-0.2\pm 0.1$  \\
$a_1(1260)^0\pi^-$ & $(6.8\pm 6.1)\times 10^{-3}$ &$\beta=+0.2\pm 0.1$ \\
$a_1(1260)^0\pi^-$ & $(5.9\pm 5.7)\times 10^{-3}$ &$\beta=-0.2\pm 0.1$ \\

$a_1(1260)^-K^0$ & $(3.1\pm 2.2)\times 10^{-7}$ & \\
$a_1(1260)^0K^-$ & $(3.8\pm 2.7)\times 10^{-7}$ & \\

$f_1(1285)\pi^-$ &  $(1.3\pm 1.2)\times 10^{-4}$ &  \\
$h_1(1170)\pi^-$ &  $(3.1\pm 1.8)\times 10^{-3}$ &  $\beta=+0.2\pm 0.1$ \\
$h_1(1170)\pi^-$ &  $(2.7\pm 2.2)\times 10^{-3}$ &  $\beta=-0.2\pm 0.1$ \\

$K_1(1270)^-\pi^0$ & $(0.8\pm 0.2)\times 10^{-6}$ & \\
$K_1(1270)^0\pi^-$ & $(1.4\pm 0.5)\times 10^{-6}$  & \\
$K_1(1400)^-\pi^0$ & $(1.1\pm 0.1)\times 10^{-6}$ & \\
$K_1(1400)^0\pi^-$ & $(2.1\pm 0.2)\times 10^{-6}$ & \\
$K_1(1270)^0K^-$   & $(13.0\pm 8.9)\times 10^{-9}$ & $\beta=+0.2\pm 0.1$ \\
$K_1(1270)^-K^0$   & $(2.8\pm 1.9)\times 10^{-9}$ & $\beta=+0.2\pm 0.1$ \\

\hline \hline
\end{tabular}
\end{center}
\end{table}

\section{Conclusions}

We have estimated branching ratios of the $\tau\to AP\nu$ decays. The axial-vector mesons $A$ 
considered are $a_1(1260)$, $f_1(1285)$, $h_1(1170)$, $K_1(1270)$, and $K_1(1400)$. The pseudoscalar 
mesons $P$ are $\pi$ and $K$. In total, 12 exclusive channels are kinematically allowed, 
first class-currents, and G-parity conserved.

To calculate the branching ratios we use a meson dominance model, where the form factors of the current
$\langle AP|J_\mu|0\rangle$ are calculated from the vector meson contributions $\rho(770)$, $\rho(1450)$,
and $K^*(982)$ with the exception of the channel $\tau^-\to f_1(1285)\pi^-\nu$, where the contribution 
is due to the axial-vector meson $a_1(1260)$. 

The channels $\tau^-\to a_1(1260)^-\pi^0\nu$, $\tau^-\to a_1(1260)^0\pi^-\nu$, and $\tau^-\to h_1(1170)\pi^-\nu$ 
decays are of order ${\cal O}(10^{-3})$. We can compare the estimated branching ratio 
$Br(\tau^-\to f_1(1285)\pi^-\nu)=(1.3\pm 1.2)\times 10^{-4}$ with the measured value 
$(3.86\pm 0.25)\times 10^{-4}$, recently reported by BABAR Collaboration \cite{babar2012}.

We observe a pattern in the branching ratios calculated in this work. The modes of order ${\cal O}(10^{-7})$ 
and ${\cal O}(10^{-9})$ are kinematically suppressed. The channels of order ${\cal O}(10^{-6})$ are Cabibbo 
suppressed and they have only the contribution of the vector meson $K^*(982)$. Finally, the channels with 
branching ratios of order ${\cal O}(10^{-3})$ and the channel $\tau^-\to f_1(1285)\pi^-\nu$ of order 
${\cal O}(10^{-4})$ are Cabibbo allowed and have two vector meson resonance contributions $\rho(770)$ and 
$\rho(1450)$, with the exception of the $\tau^-\to f_1(1285)\pi^-\nu$ where the resonance is the 
axial-vector meson $a_1(1260)$.

The processes studied in this work can contribute as intermediate states in branching ratios with four and
five pseudoscalar mesons $\pi$ and/or $K$ in final states for the $\tau$ decay modes.

Eventually, the branching ratios of order ${\cal O}(10^{-3})$ for the channels calculated in this work, 
with the exception of the channel $\tau^-\to f_1(1285)\pi^-\nu$ (which have been already measured), 
could be measured by four-prong of $\tau$ decays, with the data sample of $\tau$ lepton pairs accumulated 
by the B-factories BABAR and Belle experiments \cite{BabarBelletau}. This will not be the case for the
channels $\tau^-\to K_1(1260)\pi\nu$ and $\tau^-\to K_1(1400)\pi\nu$, with branching ratios of order
${\cal O}(10^{-6})$, in view of the sensitivity reached in some channels involving strange mesons
in final states; see Refs. \cite{babar2012, Lees2012}. Nevertheless, SuperB \cite{superB} and Belle-II 
\cite{BelleII, BelleIIa} Collaborations have programs to produce $\tau$ pairs and measure rare 
$\tau$ decays of this order.

\begin{acknowledgments}
G. C. acknowledges financial support from Promep, M\'exico. J. H. M. and C. E. V. are grateful to 
Comit\'e Central de Investigaciones (CCI) of the University of Tolima, Colombia.
\end{acknowledgments}

\end{document}